\documentclass{aa}

\usepackage[breaklinks=true]{hyperref} 
\usepackage{tablefootnote}
\usepackage{lscape}
\usepackage{afterpage}
\usepackage{xcolor}
\usepackage{booktabs}
\usepackage{longtable}
\usepackage{supertabular}  
\usepackage{caption}
\usepackage{subcaption}
\usepackage{threeparttable}
\usepackage{array}

\usepackage{array}
\newcolumntype{H}{>{\setbox0=\hbox\bgroup}l<{\egroup}@{}}
\usepackage{txfonts}
\usepackage{graphicx}

\begin{document}

   \title{The globular cluster system of the nearest Seyfert II galaxy Circinus}

   \subtitle{}

   \author{C. Obasi
          \inst{1,2}
          \and
          M. G\'omez\inst{1}
           \and
          D. Minniti\inst{1,3}
           \and
          J. Alonso-Garc\'ia\inst{4,5}
          \and
          M. Hempel\inst{1,9}
          \and
          J. B. Pullen\inst{1}
          \and
          M. D. Gregg\inst{6}
          \and
           L. D. Baravalle\inst{7}
          \and
          M.V. Alonso\inst{7,8}
            \and
          B.I. Okere\inst{2}
         }

   \institute{Instituto de Astrof\'isica, Facultad de Ciencias Exactas, Universidad Andres Bello,\\ Av. Fernandez Concha 700, Las Condes, Santiago, Chile.
              \email{c.obasi@uandresbello.edu}
         \and
             Centre for Basic Space Science, University of Nigeria, Nsukka Nigeria.
                     \and
             Vatican Observatory, V00120 Vatican City State, Italy.
                     \and
             Centro de Astronom\'{i}a (CITEVA), Universidad de Antofagasta, Av. Angamos 601, Antofagasta, Chile.
                      \and
                      Millennium Institute of Astrophysics, Nuncio Monse\~nor Sotero Sanz 100, Of. 104, Providencia, Santiago, Chile.
                      \and
                      Department of Physics, University of California, Davis, CA 95616.
                      \and
                      Instituto de Astronom\'ia Te\'orica y Experimental (IATE, CONICET-UNC),  Laprida 854, C\'ordoba, Argentina. 
                      \and
                       Observatorio Astron\'omico de C\'ordoba, Universidad Nacional de C\'ordoba, Laprida 854, C\'ordoba, Argentina. 
                       \and
                       Max-Planck Institute for Astronomy, Koenigstuhl 17, 69177 Heidelberg, Germany. 
                       }

  \date{Accepted <date> }

 
  \abstract
   {The globular cluster (GC) system of Circinus galaxy has not been probed previously partly because of the location of the galaxy at - 3.8$^\circ$ Galactic latitude which suffers severely from interstellar extinction, stellar crowding and Galactic foreground contamination. 
    However the deep near-infrared (NIR) photometry by the VISTA Variables in the Via Láctea Extended Survey (VVVX) in combination with the precise astrometry of Gaia EDR3  allow us to map GCs in this region.  
   }
   {Our long-term goal is to study and characterise the distributions of GCs and Ultra-compact dwarfs of Circinus galaxy which is the nearest Seyfert II galaxy. 
   Here we conduct the first pilot search for GCs in this galaxy. }
   {We use 
   NIR VVVX photometry in combination with Gaia EDR3 astrometric features such as astrometric excess noise
and BP/RP excess factor to build the first homogeneous catalogue  
of GCs in Circinus galaxy. A robust combination of selection criteria allows us to effectively clean interlopers from our sample.
}
   { We report the detection of$\sim$ 70 GC candidates in this galaxy at a 3 $\sigma$ 
   confidence level.  They show a bimodal 
  colour distribution with the blue peak at (G-Ks)$_0$ = 0.985$\pm$0.127 mag with a dispersion of 0.211$\pm$0.091 mag and the red peak at (G-Ks)$_0$ = 1.625$\pm$0.177 mag with a dispersion of 0.482$\pm$0.114 mag. 
  A GC specific frequency (S$_N$) of 1.3$\pm$0.2 was derived for the galaxy, and we estimated a total population of 120$\pm$40 GCs.  Based on the projected radial distribution it appears that Circinus has a different distribution of GC candidates than MW and M31.  
   }
   { We demonstrate that Circinus galaxy hosts a sizeable number of cluster candidates. This result is the first leap
towards 
understanding the evolution of old stellar clusters in this galaxy. }


   \maketitle
%

\section{Introduction}\label{Intro}

Circinus galaxy or ESO 97-G13 is a nearby spiral galaxy first observed by \cite{freeman1977large} on optical plates. The galaxy is found at J2000
R.A. = 14h 13m 09.906s, DEC = -65º 20' 20.47'' ($l$ = 311º.326, b = -3º.808), which is highly
obscured and veiled by dust and high foreground stellar density
from our Galaxy. 
The central region presents an interstellar extinction in the 
V-band of  
A$_{V}$ =  3.96 mag 
obtained by \cite{schlafly2011measuringData}.   
It has a radial velocity of 434.41 km $s^{-1}$ \cite[or a redshift of 0.001448,][]{meyer2004hipass} and it is located at about 4 Mpc away from the Milky Way (MW).
The spectroscopic observations made by \cite{oliva1994size,oliva1998spectropolarimetry}  showed 
an active nucleus confirming that Circinus is 
a Seyfert II galaxy.

The galaxy has an active core that is strong in X-ray, infrared and radio emissions \citep{Guo2019,yang2009suzaku}.  \cite{matt1996reflection} found an X-ray spectrum that is
 consistent with Compton scattering and fluorescent emission
from cold matter illuminated by an obscured active nucleus.
The IRAS fluxes of 246 Jy at 60 mm and 314 Jy at 100 mm
\citep{ghosh1992far,yamada1993search}  are consistent with a far-infrared luminosity
of 6 $\times$ 10$^9$ L$_{\odot}$ which suggest a high starburst activity. 
It has been
shown that Circinus contains a bright compact source through
radio continuum observations \citep{freeman1977large,whiteoak1985fst,
harnett1987radio,davies1998near}  and prominent
radio lobes roughly perpendicular to the galaxy's major
axis \citep{harnett1990new,elmouttie1995polarized}.  \cite{marconi1994prominent}
 detected a circum-nuclear ring in the H$\alpha$ line which implies
a high concentration of molecular gas \citep[see also][]{jones1999large,koribalski20041000}. 
 \cite{jones1999large} showed that Circinus galaxy
is over 80 arcmin long or nearly 100 kpc using its distance
of 4 Mpc from  HI observations taken with the Australia Telescope
Compact Array (ATCA). They also found
that Circinus has a large-scale spiral pattern with  the inner radius being
about 250 arcsec or 5 kpc. It contains a highly elongated
structure with non-circular motions which suggests that Circinus
is a barred spiral galaxy. Low resolution 21 cm observations
have revealed the enormous HI envelope of the galaxy,
its large-scale gas disk is shown to extend over an area more
than one degree in diameter and is argued to be much larger than
the stellar disk. The HI physical diameter of  Circinus
resembles that of the M83 galaxy \citep{jerjen2008galaxies1}.

 Circinus 
 appears to be isolated and has no known merger history.
\cite{for2012gas} presented a detailed study of the 
galaxy, investigating its star formation, dust and gas properties in
both the inner and outer disk using high-resolution Spitzer mid-infrared
images with IRAC (3.6, 5.8, 4.5, 8.0 $\mu$m) and MIPS (24
and 70 $\mu$m) and sensitive HI data from the ATCA and the 64 m
Parkes telescope. They derived A${_V}$ = 2.1 mag from the spectral distribution and
showed global star formation rates between 3 and
8 M$_{\odot}$ yr$^{-1}$.
Together with Centaurus A they are the closest active galaxies to the MW. We show in Table \ref{table1} the measured parameters of this galaxy.   
Despite all the richness and interesting features found in this
galaxy, most studies have been only limited to its nuclear
region because of the location in the sky. The 2MASS Large Galaxy Atlas \citep{jarrett20032mass}
gave us the initial view of the stellar content of Circinus, which
was limited by its dust opacity at 2 $\mu$m and its dense stellar foreground.
They have prevented a detailed investigation of its large-scale structure
and star clusters. As of today, we do not understand how the old
stellar associations of Circinus are distributed. 

Star clusters are crucial in our understanding of the evolution of galaxies. 
For this reason, we are using a combination of VISTA Variables in the Vía Láctea Extended Survey \citep[VVVX,][]{minniti2018mapping} data with Gaia EDR3 \citep{Gaia2021} 
to conduct the first search of globular clusters (GC) in Circinus galaxy. 
The paper is organized as follows: 
in Section \ref{Data} we describe the observations and in Section
\ref{sec3} we elaborate our methods to be used to search for the GCs. We
present our results in Section \ref{sec:4} and the summary and conclusions
in Section \ref{sec:5}.


\section{Observations}\label{Data}
We used the VVVX described by \cite{minniti2018mapping,obasi2021confirmation} which maps the Galactic bulge and southern disk
in the NIR with the VIRCAM (VISTA InfraRed CAMera) at
the 4.1m wide-field Visible and Infrared Survey Telescope for
Astronomy \citep[VISTA;][]{emerson2010visible} at the European
Southern Observatory (ESO) Paranal Observatory (Chile).

\begin{figure}
\begin{subfigure}[h]{0.9\linewidth}
\includegraphics[width=\linewidth]{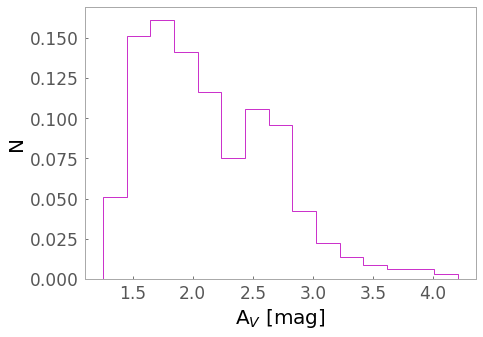}
\end{subfigure}
\caption{The total interstellar extinction distribution, A$_V$ mag, derived from the maps of \cite{schlafly2011measuringData} of all  point sources detected in the two studied VVVX tiles within the Circinus galaxy.}
\label{figA}
\end{figure}

In this work, the data used consist of the two VVVX tiles e656 and e657, which cover
the field of Circinus galaxy.  

We created a photometric catalogue using SExtractor version 2.3.1 \citep{bertin1996sextractor}, which is a program  designed to detect, deblend, measure and classify sources from astronomical images.
SExtractor is known as a useful tool to analyse overlapping objects found within the crowded region of the bulge and the disk.
In order to generate the catalogue we used SExtractor in
double-image mode with the K$_s$ (full width at half maximum, FWHM=0.5$''$) image as a reference because of
its better quality. The K$_s$ image was used for source detection,
and H and J-passband images were used for colour measurements
only.  We set our DETECT\_THRESH and ANALYSIS\_THRESH to 5 $\sigma$ in order to minimise spurious  detections.
\begin{figure*}
\includegraphics[width=\textwidth]{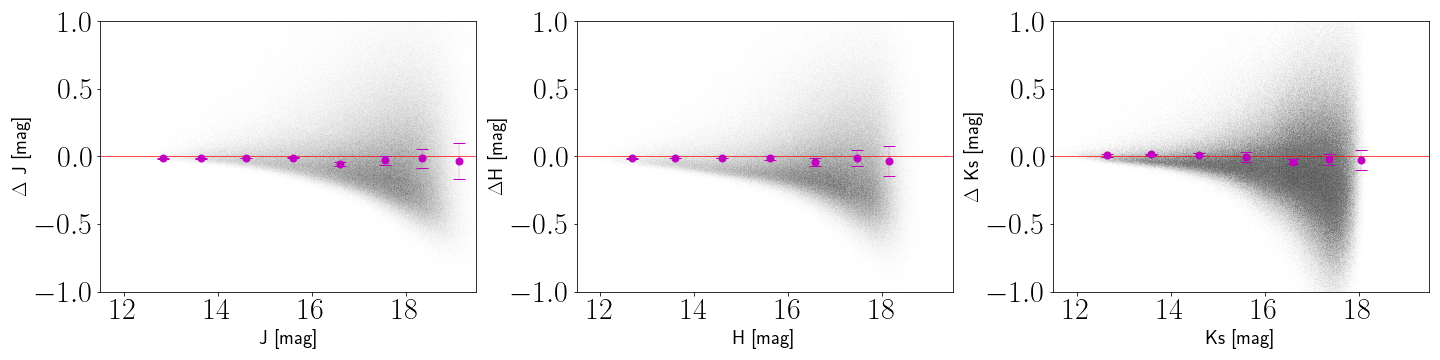}
\hfill
\includegraphics[width=\textwidth]{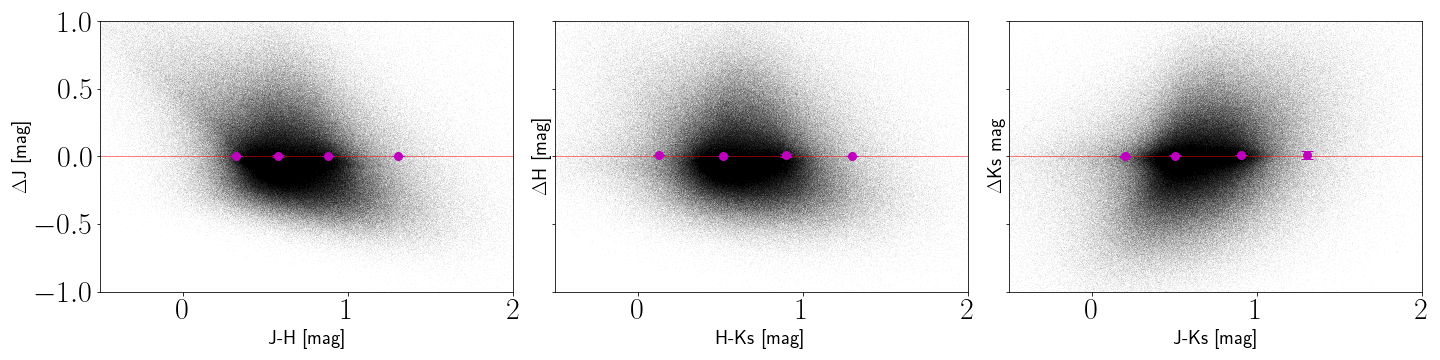}
\caption{Differences in J, H, K$_s$ as a function of magnitudes (upper panels) 
and colours (bottom panels) compared to AG22. 
 The magenta colour shows the median values of each magnitude and colour bin as well as their associated errors.}
\label{fig1}
\end{figure*}
The photometry in each of the VVVX fields e656 and e657 was
treated independently. The magnitudes were corrected for total interstellar extinctions following the maps of \cite{schlafly2011measuringData}.
Based on our detected point sources, we show in Figure \ref{figA} the total interstellar extinction distribution in the two VVVX fields which include Circinus galaxy.  The peak in A$_V$ is around 1.7 mag  with a mean value of $\sim$ 2 mag. 
The complex pattern shown by the interstellar extinction makes it necessary to perform this correction locally instead of adopting a global A$_V$. Using A$_{V}$ = 3.1 $\times$ E(B-V), we derived A$_{V}$ = 3.96 mag for Circinus centre. Relative extinctions  A$_J$= 0.280$\times$A$_V$, A$_H$=0.184$\times$A$_V$ and A$_{Ks}$=0.118$\times$A$_V$ from \cite{catelan2011vista}  were used to obtain interestellar extinctions for each  point sources.  Once the catalogues were generated for
each field, we use the point spread function (PSF) photometry to calibrate  them using a preliminary version of the new VVVX photometric catalogue (Alonso-Garc\'ia et al., in prep; hereafter AG22).  
Before the calibration, the mean differences between AG22 and our instrumental magnitudes from SExtractor (MAG\_AUTO) for the J, H and K$_s$ passbands are 0.091, 0.088, and 0.053 mag, respectively. 
In Figure~\ref{fig1}, we show the
differences in the J, H and K$_{s}$ magnitudes compared with our magnitudes after the zero point corrections (upper panels) and colours (bottom panels). 
For the upper panels, we divided our photometry into 1 magnitude bins and computed the median difference for each bin. For the bottom panels, bins of 0.4 mag were used and the median values for the differences in JHKs were also computed. 
The median values were then plotted, there seem to be no clear colour dependence. 
Our detection limits in J, H and K$_S$ passbands were 18.5, 18.0, and 17.5 mag, respectively. 
At the completion of the calibration, the two fields were concatenated as a single
catalogue. 
As there was a small superposition between both VVVX fields, duplicate sources were removed and a sample of homogeneous data for Circinus
galaxy was generated.

\section{Searching for Globular Clusters}\label{sec3}
\begin{figure*}
\centering
\includegraphics[width=0.8\textwidth,height=0.50\textwidth]{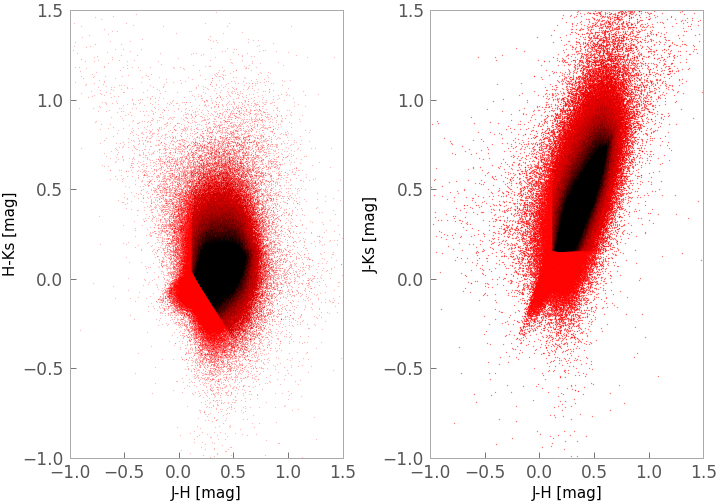}
\hfill
\includegraphics[width=0.8\textwidth,height=0.50\textwidth]{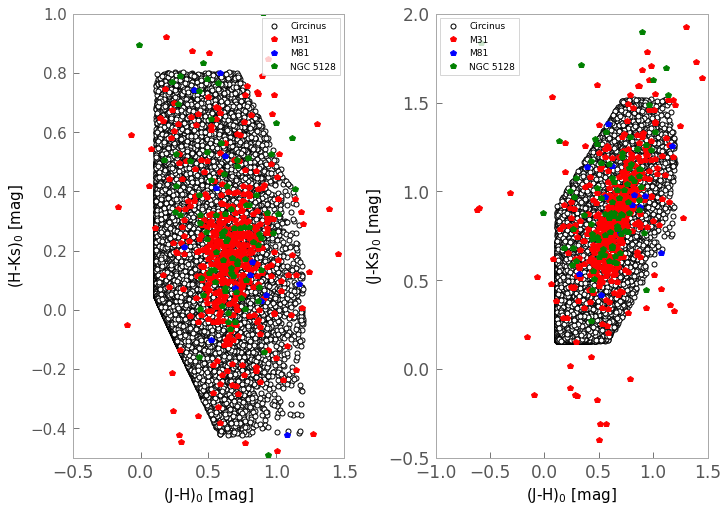}
\caption{The colour-colour diagrams H-K${_s}$ vs J-H  and J-K${_s}$ vs J-H (upper panels) showing in black the colour cuts used for the analysis. Bottom panels reproduce the same region above, now overplotted with GCs from M31 \citep{wang2014new}, M 81 \citep{M81nantais2012vizier} and NGC 5128 \citep{woodley2009ages,taylor2015observational}}
\label{fig0}
\end{figure*}

First, detected objects were classified into point and extended sources using the CLASS$\_$STAR index provided by SExtractor. The CLASS$\_$STAR is a stellarity index that is associated with the light distribution of the source.  
It  ranges from 1 for point sources such as stars to 0 for extended objects like galaxies.
To disentangle the different source types, SExtractor  relies on a multi-layer feed-forward neural network trained using supervised learning to estimate a posteriori probability that the detection would be either a point source or an extended object. We selected point sources by adopting CLASS$\_$STAR $\geq$ 0.5. 902,667 objects passed this selection which represents about 80$\%$ of all source detected.  We then applied colour cuts in order to photometrically select those points sources that have colours and magnitudes consistent with those expected for GCs in other galaxies. From the catalogue of 
\cite{wang2014new} there are 568 spectroscopically confirmed GCs  
\citep{galleti20042mass}\footnote{This is the 2012 revised and updated Bologna catalogue of M31 GCs and candidates. It contains 625  confirmed GCs, 568 of which have a match in \cite{wang2014new} catalogue.\\\url{https://cdsarc.cds.unistra.fr/viz-bin/cat/V/143}} that we used for our colour selection. Our adopted colour cuts were restricted to $\pm$2 $\sigma$ statistical level from the mean GCs colours. This allows us to minimise the contamination from foreground stars. In Figure \ref{figB}, we show the complete \cite{wang2014new} GC catalogue in blue and in magenta the confirmed GCs.
The mean colour obtained and their limits as applied to our selection are given as 0.655, 0.113  $\leq$ (J-H)$_0$ $\leq$ 1.197 mag, 0.189, -0.424$\leq$ (H-K$_{s}$)$_0$  $\leq$ 0.802 mag and 0.839, 0.159 $\leq$ (J-K$_{s}$)$_0$ $\leq$  1.519 mag. We obtained 791662 sources after the colour cuts, which account for 87.7$\%$ of 
the total sources. The colour-colour selected region used for the rest of the analysis is shown in Figure~\ref{fig0}. Upper panels show our selected region with black points and the bottom panels our selection overlaid with confirmed
GCs of M31 \citep{wang2014new}, M81 \citep{M81nantais2012vizier} and NGC5128 \citep{woodley2009ages,taylor2015observational}.  They indicate that GCs occupy a wide range of colours.
 

Additionally, the sources that satisfy our colour cuts
were cross-matched with Gaia EDR3 within a radius of  0.5$''$. 
Gaia is an all-sky astrometric catalogue of more than a billion sources. The spatial resolution of the survey has improved from 0.40$''$ in Gaia DR2 \citep{Lindegren2018} to 0.18$''$ in Gaia EDR3 \citep{Gaia2021}  GCs have effective radii of $\sim$ 2-10 pc as given in \cite{harris1991globular}. At the distance of Circinus, they appear as marginally extended sources to Gaia which has a resolution of 0.059$''$ per pixel in the scanning direction. By combining the VVVX data with a pixel size of 0.34$''$ and Gaia EDR3 data, we can take advantage of some of the Gaia astrometric features to resolve the GCs of Circinus galaxy. The Gaia EDR3 catalogue for a circular region of radius 1.6 deg. centred on RA = 213.29$^{\circ}$ and
DEC = -65.33 $^{\circ}$ (corresponding to a diameter of 108 kpc at the distance of Circinus)
was downloaded. At low Galactic latitudes the contamination by foreground stars is very severe \citep{baravalle2018searching}. Because Circinus is located just at b = -3.8$^{\circ}$, below the Galactic plane, and also veiled by the Galactic disk, most of the detected objects are therefore expected to be stars within the Galactic plane. The use of Gaia EDR3 is crucial in eliminating MW foreground stars using their proper motions and parallaxes.
Any source with a significant proper motion or parallax can be excluded from our analysis as being a foreground object. The total proper motion of each
object was obtained assuming that the combined proper motion has a $\chi^{2}$
distribution with two degrees of freedom. In this way, the
proper motion likelihood of 1-cumulative distribution function (CDF) of the
Rayleigh distribution was assigned to each
object.  In this study, we considered only
objects whose proper motion probability values are consistent with zero at
2$\sigma$.  Sources with probability values smaller than  -1 or greater than 1 are considered as foreground stars, and were assigned proper motion likelihood = 0 and therefore were removed from  our analysis.
    
A similar procedure was adopted for the 
the parallax. We assumed that the measurements follow a Gaussian distribution. If the parallax is significant at 2$\sigma$, the likelihood assigned is -0.4 $\leq$ Plx $\leq$ 0.4 
Sources with $>$ 3$\sigma$ parallax measurements were assigned parallax likelihood = 0 and were not considered in the rest of our analysis.
As the fourth step, the sources that satisfied our previous criteria were then subjected to some tests with the astrometric excess noise (AEN) and phot$\_$bp$\_$rp excess factor (BRexcess).
These are two Gaia parameters that are found very useful in disentangling point and extended sources. \cite{voggel2020gaia}  demonstrated how these two parameters might be used to select extended sources such as GCs out to a distance of 25 Mpc. The AEN is a statistical computation that quantifies the goodness of the fit of their five parameter astrometric model to the astrometry for each target. The AEN is measured in milli-arcseconds (mas) and is equal to 0 for a well fit star and larger for a poorer fit. It has been shown in \cite{voggel2020gaia} that objects that are extended sources generally have larger AEN values and this could be exploited in selecting marginally resolved objects such as GCs. On the other hand, the BRexcess gives the ratio of the sum of the flux of the Blue photometer (BP, 3300-6800 $\r{A}$) plus the Red photometer (RP, 6400-10500 $\r{A}$) with the flux in the broad- G passband. Its importance and usefulness come from the unique way these fluxes are derived. Both the BP and RP magnitudes are measured from the flux within a larger aperture of 3.4$\times$2.1 arcsec$^{2}$, while the G magnitude is obtained from profile fitting with an effective resolution of 0.4$''$  \citep{evans2018gaia}. The variation means that extended sources have a higher BRexcess than point sources, since fitting an effective point-source profile to an extended source in the G passband misses some of the total light. \cite{voggel2020gaia} used the combination of these two parameters to identify over 600 new luminous cluster candidates in the halo of NGC5128 out to a projected radius of 150 kpc. In a follow-up study by \cite{hughes2021ngc}, they used this method to estimate a total population of 1450$\pm$160 GCs to a projected radius of 150 kpc in NGC5128 
demonstrating its robustness. In using both the AEN and BRexcess they calculated the likelihood that a source is an extended source (GC) within a radial range of 10-60$'$ from the galaxy center to define a line of 3$\sigma$ above the mean foreground star AEN and BR value derived are given in equations \ref{equ:1} and \ref{equ:2}.

\begin{equation}\label{equ:1}
AEN_{3\sigma}=0.297+5.63 \times10^{-8}e^{0.895G}
\end{equation}
\begin{equation}\label{equ:2}
BRexcess_{3\sigma}= 1.26+2.79 \times 10^{-8}e^{0.853G}.
\end{equation}

\begin{figure*}
\centering
\includegraphics[width=\textwidth]{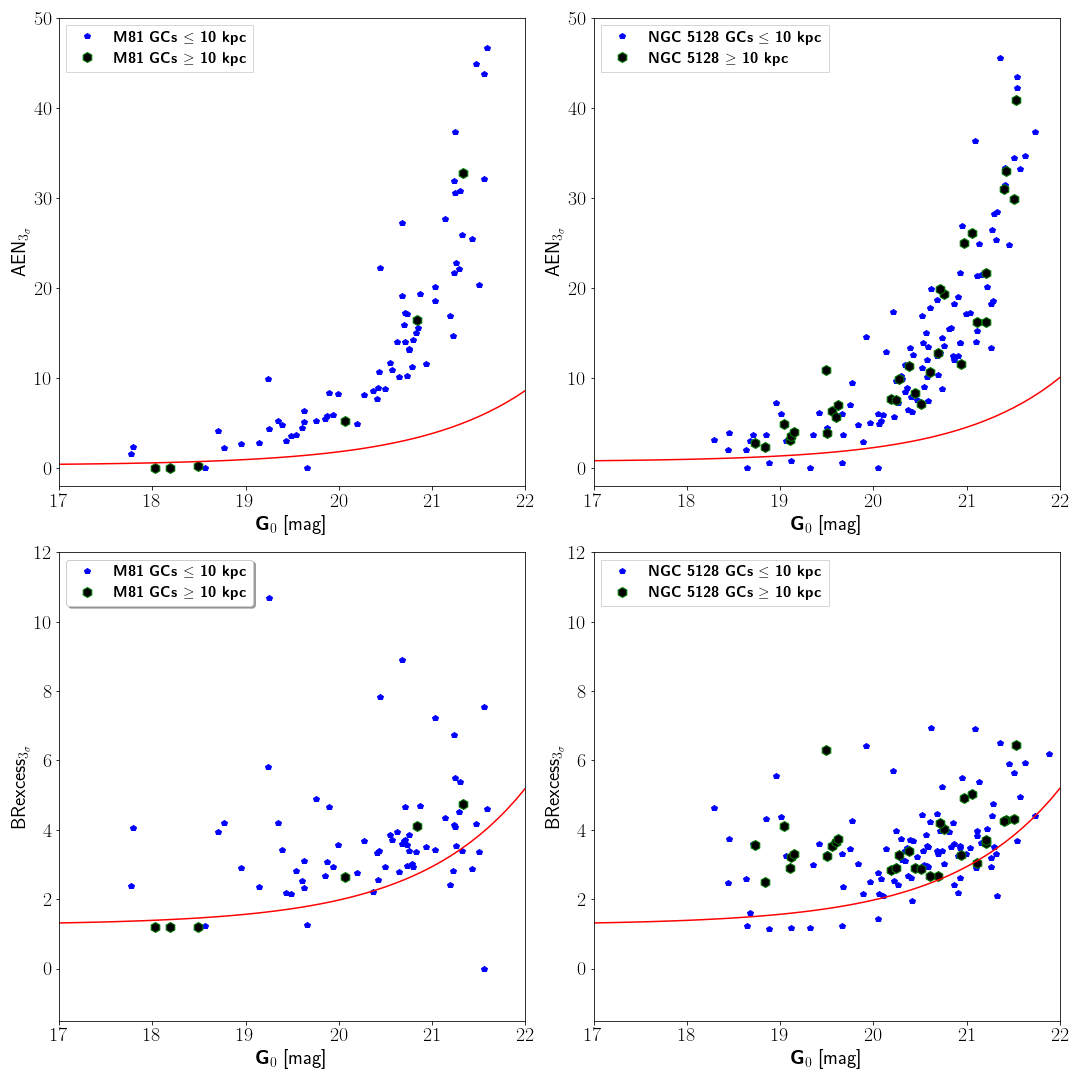}
\caption{Spectroscopically confirmed GCs in M81  \citep{M81nantais2012vizier} and NGC5128  \citep{woodley2009ages} combined with \cite{taylor2015observational} and Gaia catalogues.  
The AENs and BRexcess as a function of the G magnitudes are shown in the upper and bottom panels, respectively. Blue points represent GCs at $\leq$ 10 kpc from the galaxy centre while  black points, those GCs at $\geq$ 10 kpc. The red curve defines 3$\sigma$ above the mean foreground stars. 
}
\label{fig2}
\end{figure*}

We tested the above procedure by using
spectroscopically confirmed GCs in M81 of \cite{M81nantais2012vizier} and NGC5128 of \cite{woodley2009ages} and \cite{taylor2015observational}. The M81 catalogue comprises 108 confirmed GCs and after  cross matching with the Gaia EDR3 catalogue we found 77 counterparts. We adopted a distance modulus of 27.64 $\pm$ 0.09 for M81  \citep{freedman1988distances} and corrected for relative extinctions in G passband with the relation A$_G$=0.86$\times$A$_V$ given by \cite{fernandez2021pristine}. For each source, the distance to the centre was computed and six sources had distances greater than 10 kpc from the galaxy centre. For NGC5128 we used two catalogue of spectroscopic confirmed catalogues comprising 37 GCs by \cite{woodley2009ages}  and 140 GCs by \cite{taylor2015observational}. 
These catalogues were then cross matched with Gaia EDR3 and 134 sources were found in common. For NGC5128, a distance modulus of 27.9$\pm$ 0.2 was adopted from \cite{rejkuba2004distanceNGC5128}.  
Then we proceed to compute the distance of each GC from the centre and a total of 31 sources have radial distances greater than 10 kpc. Figure~\ref{fig2} shows the properties for NGC5128 and M81 using equations \ref{equ:1} and \ref{equ:2} of \cite{hughes2021ngc} as AEN vs G magnitudes and BRexcess vs G magnitudes (upper and lower panels, respectively).  Blue points are sources with radial distances less than 10 kpc while black points are sources with distances greater than 10 kpc. The red curve represents the 3$\sigma$ probability that a source is more extended than a point source (in our case GCs). 
We find the brighter objects tends to have smaller AEN values, which could be attributed to the statistical uncertainties in Gaia observations.  We might conclude that all 
the black points are consistent with being
point sources for NGC5128.  
Three points are below the curve for M81 and they might be most probably misclassified foreground stars \citep[see ][]{hughes2021ngc}. It's important to note that the above procedure is sensitive to objects with distances $>$ 10 kpc to the galaxy centre. In this analysis, we only considered the outermost parts of the galaxy where the internal extinction is less severe compared to the central part.
We applied this procedure to all the sources in  Circinus galaxy that satisfied our previous criteria. There are 260 sources that satisfied the AEN criterion and 176 sources the BRexcess cut. 
As additional  
step, the sources should also satisfy 
the spread model parameter  criterion. $\Phi$, is another star-galaxy classifier that was added in the newest SExtractor versions. 
The value of this parameter is usually zero for stars, positive for extended sources and negative for detections smaller than the PSF, such as cosmic rays, etc . 
We adopted the same cut used by  \cite{baravalle2018searching}  for searching extragalactic sources in VVV data ($\Phi$ > 0.002). 
For Circinus galaxy, 141 of the sources satisfy all the above criteria.

Finally, we also examined the ellipticity and FWHM of the sources inspecting their morphology (roundness and size) as one critical criterion for selecting GCs \citep[see][]{minniti1996globular,
alonso1997infrared,rejkuba2001deep}. Both features are convenient parameters to obtain a bonafide GCs sample. However, using only FWHM as discriminating criterion will lead to the rejection of both very concentrated and very loose clusters. Conversely using only roundness will underestimate the GCs sample. By applying these two stringent criteria together, we can trade off  masquerading interlopers as well as some true genuine GCs. We adopted  
$\varepsilon$ $\leq$ 0.4 consistent with \cite{georgiev2009globular} and \cite{woodley2010globular} and 2$\leq$FWHM (pixels) $\leq$ 4 consistent with \cite{alonso1997infrared,rejkuba2001deep}  obtaining a final sample of 78 potential Circinus GCs within the colour limits. 

\begin{figure*}
\centering
\includegraphics[width=\textwidth]{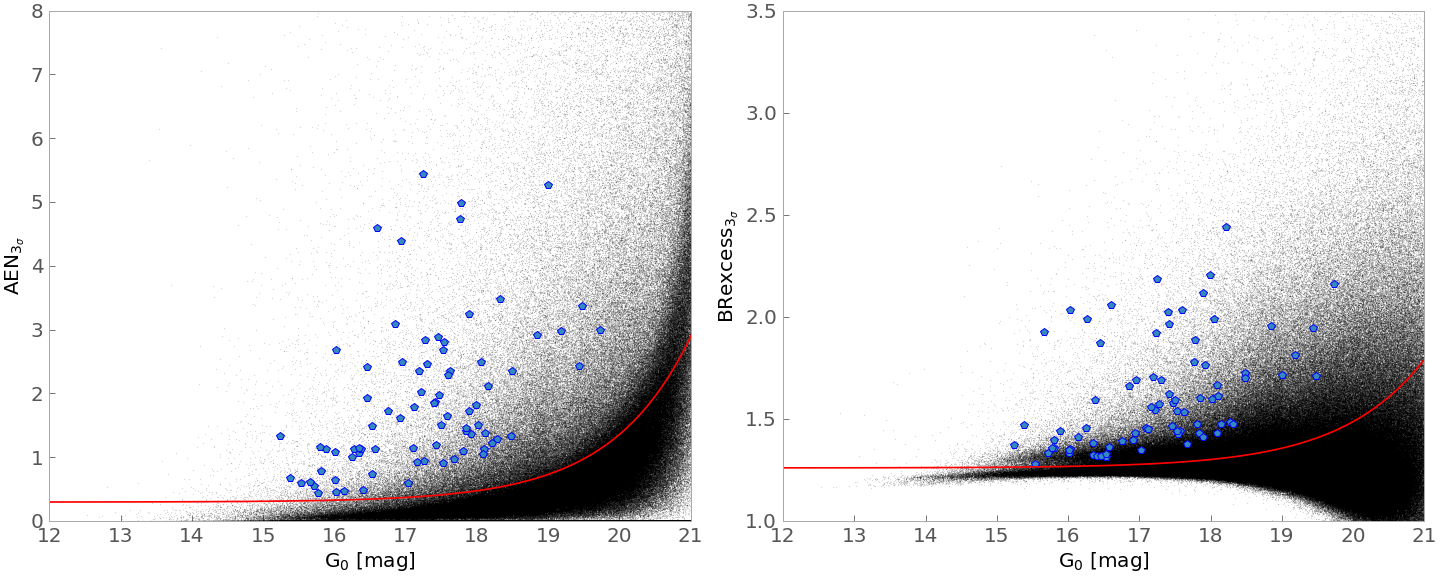}
\caption{
AENs and BRexcess as a function of the G magnitudes (left and right panels, respectively).  The point sources are over-plotted with our final sample in blue.  The red curves  show the 3$\sigma$  criteria  derived by \cite{hughes2021ngc}  to select GCs in NGC 5128.}
\label{fig3}
\end{figure*}

In Figure~\ref{fig3}, we show the AEN as a function of G magnitudes with equation \ref{equ:1} (left panel) and BRexcess as a function of G magnitudes together with equation \ref{equ:2} (right panel) for the 902,667 points in Circinus galaxy. Over plotted in blue are our final GC candidates.  

\begin{table*}[!th]

\centering
\begin{threeparttable}[b]
\caption{Parameters of  Circinus galaxy.}
\label{table1}
\begin{tabular}{@{}|l|l|l|l|@{}}
\toprule
\midrule
& &         & Reference   

\\
Central position        & $\alpha$,$\delta$(J2000) & 14$^{h}$13$^{m}$10$^{s}$.2, -65$^{\circ}$ 20$'$ 20$''$         & 1     \\

                        & $l$,$b$        & 311$^{\circ}$.3,-3$^{\circ}$.8                        & 1,3    \\
Type                    &     &Sb                                  & 1      \\
Distance                & D          & 4.0$\pm$0.8 Mpc                & 1,2       \\
                        &            &                                &              \\
Optical extent at B$_{25}$ &     & 6.$'$9$\times$2.$'$7                & 1      \\
H I extent**             &            & 32$'$$\times$                15$'$    & 2      \\
Position angle          & PA$_{H I}$        & 210$^{\circ}$ $\pm$5$^{\circ}$                            & 2      \\
Inclination angle       & i$_{H I}$        & 65$^{\circ}$ $\pm$2$^{\circ}$                             & 1      \\
Systemic velocity       & $\upsilon$$_{sys}$       & 439$\pm$2 km $s^{-1}$ (H I)                  & 1 \\
Heliocentric velocity     &            & 380$\pm$47 km s$^{-1}$ (optical)              &1        \\
Velocity width          & $\Delta$$_{\upsilon 20}$  & 290$\pm$10 km s$^{-1}$(H I)                    & 1      \\
Rotation velocity       & $\upsilon$rot       & 152$\pm$7 km s$^{-1}$(H I)                     &   1     \\
Far-infrared luminosity & LFIR       & 6$\times$ 10$^{9}$ L$_{\odot}$        & 1  \\
H I mass                 & M$_{ HI}$        & 7.2$\pm$0.5$\times$10$^{9}$ M$_{\odot} $                   & 2      \\
                        &            & 9$\times$10$^{9}$  M$_{\odot} $                          & 3 \\
Total mass              & M$_{tot}$       & 1.2$\pm$0.2$\times$10$^{11}$ M$_{\odot}$ & 1      \\
                       
Global star formation rate  & SFR        & 3 - 8 M$_{\odot}$ yr$^{-1}$  &  3 \\
 \bottomrule
\end{tabular}

\begin{tablenotes}
    \item **The half width of the H I column density distribution after Gaussian-correction.
    \item References: [1] \cite*{jones1999large}, [2] \cite{freeman1977large}, [3] \cite{for2012gas}
  \end{tablenotes}
 \end{threeparttable}

\end{table*}


\section{Results}\label{sec:4}

\begin{figure*}
\centering
\includegraphics[width=\textwidth]{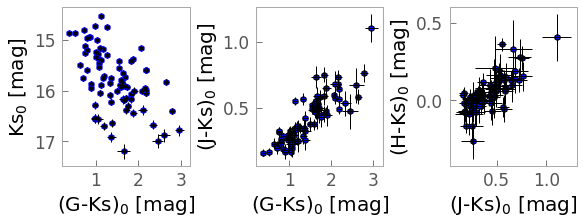}
\caption{NIR-optical colour-magnitude diagram (left panel), NIR-optical colour-colour diagram (middle panel) and NIR colour-colour diagram (right panel)
for the GC candidates in Circinus galaxy.
}
\label{fig5a}
\end{figure*}

\begin{figure*}
    \centering
    \includegraphics[width=10.cm]{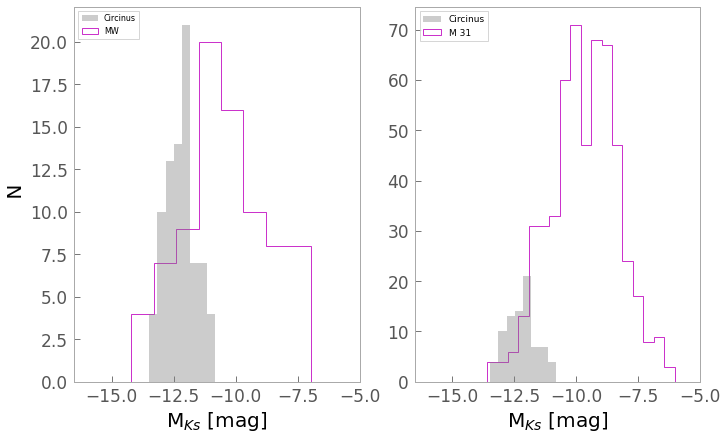}
    \caption{
    Luminosity function of the Circinus GC candidates in the K$_s$ passband together with that of the MW \citep{nantais2006nearby} and M31 \citep{wang2014new}  (from left to right). }
    \label{fig:7aa}
\end{figure*}

Our final sample is composed of 78 GC candidates (including contaminators) found in Circinus galaxy. We corrected for interstellar extinction in each of the objects.
Figure \ref{fig5a} shows the (G-K${_s}$)$_0$ vs K${_s}$ colour-magnitude diagram and the
(G-K${_s}$)$_0$ vs (J-K${_s}$)$_0$, (G-K${_s}$)$_0$ vs (H-K${_s}$)$_0$ colour-colour diagrams (left, middle and right panels). 
In Figure \ref{fig:7aa} we show the NIR luminosity function (M$_{Ks}$) of the 78 GCs compared with those of the MW   \citep{nantais2006nearby} and M31 \citep{wang2014new}. The MW sample contains 99 GCs while M31 have 568 confirmed GCs. The Figure shows that we are only detecting the brightest GCs in Circinus.   
Our sample did not reach the expected GCs turnover 
in infrared of M$_{ks}$=-10 mag, (if we assume (V-Ks) mean colour of 2.5 mag), it is difficult  to accurately estimate the total number of GCs (N$_{GC}$) in Circinus and  their specific frequency (S$_N$) which defines N$_{GC}$ per unit galaxy luminosity normalised to a galaxy with an absolute V magnitude of -15  \citep{harris1981globular}. However, we can  use the relation given in \cite{harris2013catalog} to estimate roughly N$_{GC}$. The relation reads \begin{equation}
  N_{GC}= ( 300\pm 35)\left[(\frac{R_e}{10 kpc})(\frac{\sigma_ e}{100 km/s})\right]^{1.29\pm0.03}
\end{equation}
where R$_e$ is the effective radius of the galaxy and $\sigma$$_e$ is the velocity dispersion. The above parameters for Circinus have been reported where R$_e$=6.5kpc$\pm$1kpc \cite{jarrett20032mass} and $\sigma$ $_e$=75$\pm$20km/s \cite{hu2008black}. The total number of GCs in Circinus is therefore estimated to be 120$\pm$ 40. With this result, we can calculate the S$_{N}$ of GCs  using the equation provided in  \citep{harris1981globular} 
represented by 
\begin{equation}\label{equ:3}
    S_{N}=N_{GC} 10^{0.4(M_V^T+15)}
\end{equation}  where M$_V^T$ is the absolute visual magnitude. Given that the apparent magnitude is 12.1 mag \citep{de1991book} and
A$_V$ = 3.96 mag, this implies that the absolute magnitude
M$_V$ =-19.87 mag, we obtained S$_N$ of 1.3$\pm$0.24. This result is consistent with S$_N$ values for spiral galaxies usually $\le$ 1 \citep{van1982globular}.

We tested for bimodality in the colour distribution of our sample. Most GCs found in massive galaxies have shown to exhibit bimodal colour distributions \cite[see][]{brodie2006extragalactic,brodie2014sages,harris2017globular}. Nonetheless, in some studies \cite{gebhardt1999globular,chies2012optical,beasley2018single}, there have been reports of uni-modal colour distributions for GCs which were argued to have formed at the same time in isolated galaxies with no history of mergers. We used the Gaussian mixture model (GMM)\footnote{GMM code employs the likelihood-ratio test to compare the goodness of fit for a double Gaussian vs a single Gaussian. It estimates the probability of a best-fit double model providing the means,widths of the two components, their separation DD in terms of combined widths and their overall kurtosis. Importantly, GMM also provides the positions, relative widths and fraction of objects associated with each peak as well as their uncertainties based on bootstrap resampling.} 

\begin{figure*}
    \centering
    \includegraphics[width=8.cm]{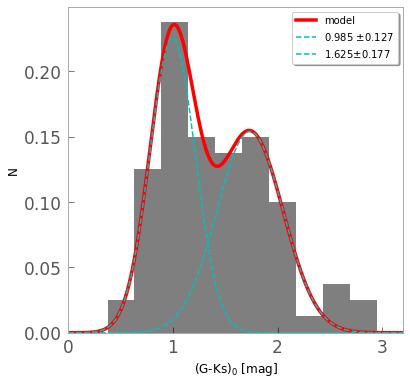}
    \caption{The (G-Ks)$_0$ colour distribution of our GC candidates .}
    \label{fig:fig26}
\end{figure*}

\begin{table*}[]
\centering
\caption{GMM test results}
\label{table3}
\begin{tabular}{@{}llllllllll@{}}
\toprule
 & Colour & N & Blue & Red & f$_{blue}$ & f$_{red}$ & Kurt & DD & p($\chi$$^2$) \\ \midrule
 & (G-J)$_0$ & 78 & 0.964$\pm$0.125 & 1.950$\pm$0.337 & 56.8$\pm$19.3 & 21.2$\pm$19.3 & 0.594 & 3.17$\pm$1.17 & 0.037 \\
 & (G-H)$_0$ & 78 & 1.095$\pm$0.172 & 1.771$\pm$0.315 & 47.5$\pm$22.0 & 30.5$\pm$22.0 & 0.339 & 2.58$\pm$0.92 & 0.285 \\
 & (G-K${_s}$)$_0$ & 78 & 0.985$\pm$0.127 & 1.625$\pm$0.308 & 33.8$\pm$16.4 & 44.2$\pm$16.4 & 0.308 & 2.08$\pm$0.69 & 0.039 \\ \bottomrule
\end{tabular}
\end{table*}


developed by \cite{muratov2010modeling}  
specifically to test for bimodality in GC systems. The model shows that our sample is well described by a  bimodal distribution with the blue peak at (G-K${_s}$)$_0$ = 0.985$\pm$0.127 mag with a dispersion of 0.211$\pm$0.091 mag and the red peak at (G-K${_s}$)$_0$ = 1.625$\pm$0.0177 mag with a dispersion of 0.482$\pm$0.114 mag.  A similar bimodal distribution can also be observed in other colours combinations as shown in table \ref{table3} where we present the results of the GMM analysis for the colours (G-J)$_0$, (G-H)$_0$, and (G-${Ks}$)$_0$. In column one, we present the colour, in column 2 the number of objects, in columns 3 to 4 the blue and red peaks, in columns 5 to 6, we provide the fraction of objects in each peak, and in columns 7 to 9, we show the Kurtosis, and the peak separations (as measured in units of the fitted $\sigma$$_s$ for the two Gaussian models) as well as the P($\chi$ ${^2}$) values. Two of the three colours that have been inspected show consistency between the blue and red fractions in the colour distribution. The third colour has a bias favoring the red fraction. Nevertheless, the low P($\chi$ ${^2}$) values from the GMM tests confirm that a bimodal Gaussian distribution is preferred over a unimodal distribution.  
Figure~\ref{fig:fig26} shows this colour distribution of
GC candidates with a bin size of 0.1 mag, giving another evidence of their nature. The red curve shows the fit. 

\begin{figure*}
\centering
\includegraphics[width=\textwidth]{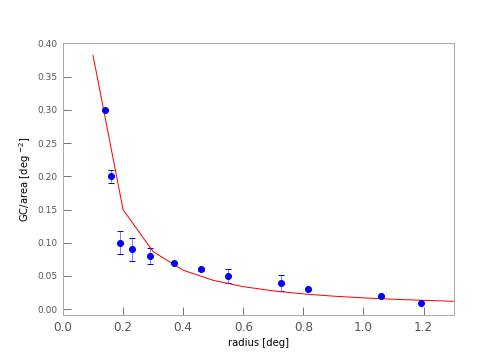}
\caption{The surface density profile of our GC  candidates in radial bins. The surface density follows a power law distribution with $\rho_{GC}$ $\propto$r$^{-1.4}$
represented by the red solid curve.}
\label{fig8a}
\end{figure*}

Also, the density distribution follows a power law which is frequently observed for GCs with $\rho_{GC}$$\propto$r$^{-1.4}$.   
 The number of GCs is still uncertain for determining  a proper radial profile. However, it is clear from Figure~\ref{fig8a} that our cluster candidates follow a distribution which closely resembles many other extragalactic GC systems \cite[see e.g ][]{gomez2004globular,brodie2006extragalactic}. Even if the true profile and background level remains to be determined, a clear concentration in the central parts of Circinus can be seen, which favours a true GC nature of our sample. 
 
\begin{figure*}
    \centering
    \includegraphics[width=8.cm]{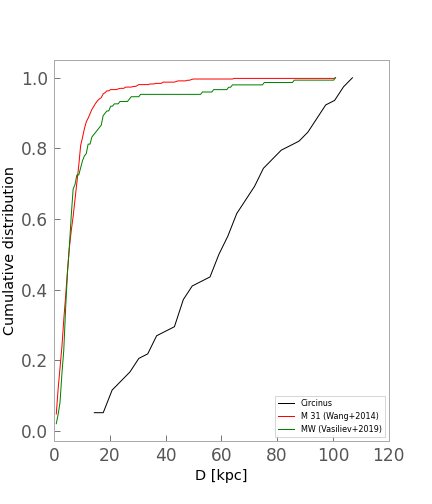}
    \caption{The cumulative distribution of GCs with increasing galactocentric distance D for Circinus, M31 and the MW in black, red and green curves, respectively.   
    }
    \label{fig:fig27}
\end{figure*}

\begin{figure*}
\centering
\includegraphics[width=\textwidth]{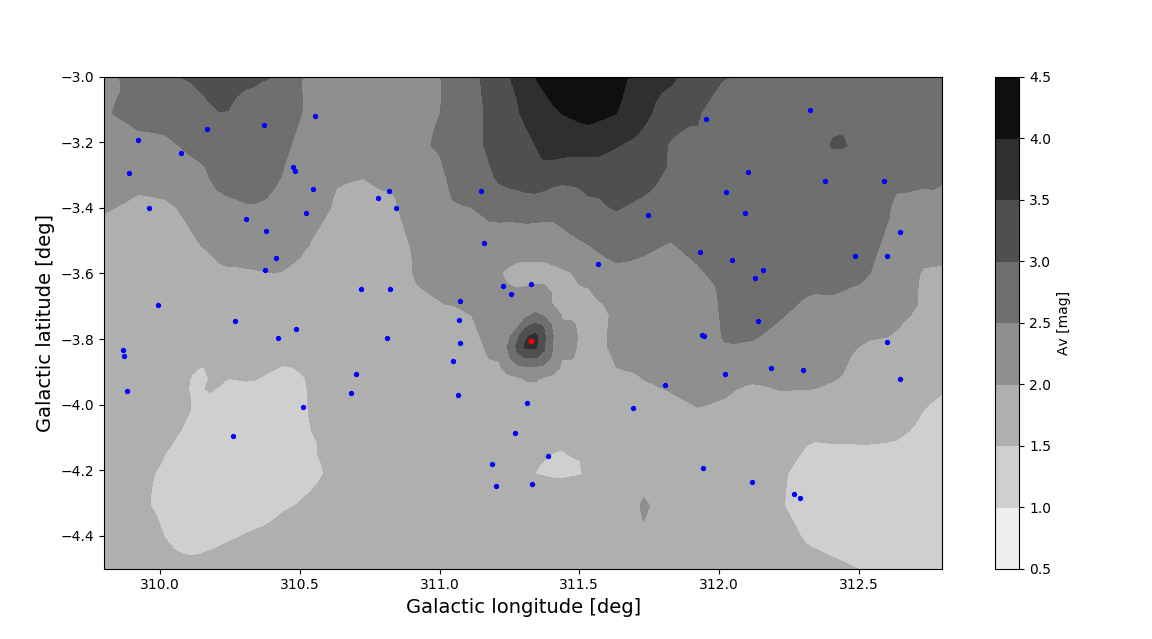}
\caption{The spatial distribution  of the 78 GC candidates in the Circinus galaxy.  The candidates are represented 
with blue points. 
The red dot indicates the adopted center of Circinus.
The total optical $A_V$ insterstellar extinctions  
from the  maps of \cite{schlafly2011measuringData} are superimposed in a grey gradient with the bar scale also shown. 
}
\label{fig8}
\end{figure*}

In Figure~\ref{fig:fig27} we compare the projected radial distances of Circinus, M 31 and MW GCs. As can be observed, large concentrations of clusters are found for D$<$ 30 kpc in M 31 and MW. We used the 568 confirmed GCs in  M31 \citep{wang2014new} and 150 in the MW catalogue \citep{MWvasiliev2019vizier} for this comparison of which 93.3$\%$ and 81.3$\%$ are within distances $\leq$ 30 kpc compared to 12.5$\%$ for Circinus. In light of these findings, a fundamental question arises regarding the level of contamination in our sample, as well as the similarity in GC distributions between MW, M31, and Circinus. As part of our effort to provide some insight into this question, we examined two empty fields at opposite ends of Circinus that were about 2.4 degrees away from Circinus. 
The fields are located at similar latitudes and their extinction patterns are relatively similar to those of  Circinus with a mean A$_V$ $\sim$ 2 mag. 
As far as the area covered is concerned, it is the same by construction. We applied the exact same procedure we used for analysing Circinus. In each of the fields, we found 8 and 10 sources that met our adopted criteria.  These amount to 10.3\% and 12.8\% contamination respectively.  
Therefore, we estimate that our sample is contaminated with 11.5$\%$ interlopers. Based on the projected radial distribution it appears that Circinus has a different distribution
of GC candidates than MW and M31.
However, the data available at present do not allow a conclusive assessment of the differences in GC radial distribution. Our sample must be cleaned up to eliminate interlopers, as well as searched for GCs in the bulge and disk to get a good idea of the overall distribution before such claim can be made. Follow-up spectroscopic observations of these candidates will help to weed out contaminators. The ratio of our 70 GC candidates (interlopers accounted for) with the estimated 120$\pm$40 GC
population expected in Circinus, suggest that about 58$\%$ of the total GC population is located in the
halo.  
In Figure~\ref{fig8}, we show the spatial distribution of the 78 selected  GC candidates in Galactic coordinates. The regions with the same optical total interestellar extinction A${_V}$
derived from the extinction maps of \cite{schlafly2011measuringData} are superimposed with different A$_V$ levels 
as shown with the grey-bar.  
The closest object to the galaxy centre is located at a projected radial distance of about 11 kpc. About 80$\%$ of the cluster candidates are located in regions with low A$_V$ interestellar extinctions (0.5 to 2.0 mag). However, more GC candidates might be missing at central regions with higher extinctions. 
Table \ref{table2} shows the 78 GC candidates, listing the identification in column (1), the J2000 equatorial coordinates in columns (2) and (3), the J$_0$, H$_0$, and K$_{s}$ $_0$ magnitudes with their respective errors in columns (4) to (9), the Gaia magnitude in column (10),
the FWHM in column (11), the ellipticity and the A$_V$, A$_J$, A$_H$, A$_{Ks}$, A$_G$ extinctions in columns (12) to (17) and the galactocentric distances in column (18).


\section{Conclusions}\label{sec:5}
This is the first study of the old stellar association in Circinus galaxy which is the nearest known Seyfert II galaxy. 
We performed our search using VVVX photometry together with the Gaia EDR3 catalogue. Taking advantage of the sharp contrast and reduced reddening sensitivity in the NIR VVVX frames and the excellent resolution of the Gaia EDR3 astrometry features (AEN and BRexcess), 
we found a total of 78 GC candidates that satisfied all our stringent criteria. 
The GC colour distribution is well described by two normal functions with a blue peak at (G-Ks)$_0$ = 0.985$\pm$0.127 mag with a dispersion 0.211$\pm$0.091 mag and a red peak at (G-Ks)$_0$ = 1.625$\pm$0.177 mag with a dispersion of 0.482$\pm$0.114 mag. We estimated N$_{GCs}$ of 120 $\pm$ 40 and derived an S$_N$
of 1.3 $\pm$ 0.2
The density distribution follows a power law with $\rho_{GC}$ $\propto$r$^{-1.4}$. The most distant GCs that we detected are located at a projected distance of about 107 kpc from the center of the galaxy, indicating that Circinus has an extended GC system. The true nature of these objects must be determined through follow-up spectroscopic observations.

  

\begin{acknowledgements}

We gratefully acknowledge the use of data from the ESO Public Survey program IDs 179.B-2002 and 198.B-2004 taken with the VISTA telescope and data products from the Cambridge Astronomical Survey Unit.

This work has made use of data from the European Space Agency (ESA) mission
{\it Gaia} (\url{https://www.cosmos.esa.int/gaia}), processed by the {\it Gaia}
Data Processing and Analysis Consortium (DPAC,
\url{https://www.cosmos.esa.int/web/gaia/dpac/consortium}). Funding for the DPAC
has been provided by national institutions, in particular, the institutions
participating in the {\it Gaia} Multilateral Agreement.  We also acknowledge the comments of
the anonymous reviewer whose positive feedback helped to improve the quality of this paper 
 D.M. gratefully acknowledges support by the ANID BASAL projects ACE210002 and FB210003. 
D.M. and M. G. also gratefully acknowledge support by Fondecyt Project No. 1220724.
 J.A.-G. acknowledges support from Fondecyt Regular 1201490 and from ANID – Millennium Science Initiative Program – ICN12\_009 awarded to the Millennium Institute of Astrophysics MAS.

\end{acknowledgements}
-----------------------------------------

 \bibliographystyle{aa} 
   \bibliography{Ref2.bib} 
  
\begin{appendix}
\section{Colour cuts \& Table Listing GC candidates}

\begin{figure*}
\centering
\includegraphics[width=\textwidth]{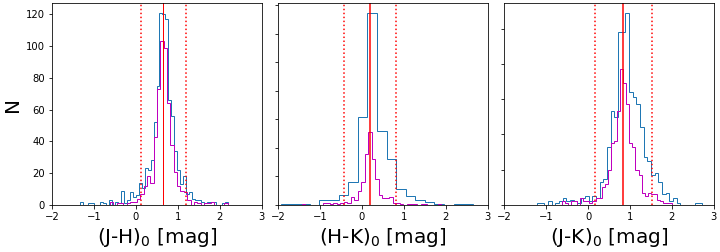}
\caption{2$\sigma$ colour cuts (broken red vertical lines) with respect to the peak colour distribution of the confirmed GCs catalogue of \citep{wang2014new} in M31. The magenta colour shows the distribution of the the confirmed Gcs in the catalogue. }
\label{figB}
\end{figure*}

\begin{landscape}
\onecolumn
\begin{longtable}{@{}llllllllllllllllll@{}}
\caption{NIR VVVX Photometry, colour and shape parameters of our cluster candidates.}
\label{table2}\\
\toprule
ID &
  \begin{tabular}[c]{@{}l@{}}RA\\ DD\end{tabular} &
  \begin{tabular}[c]{@{}l@{}}DEC\\ DD\end{tabular} &
  \begin{tabular}[c]{@{}l@{}}J\\ mag\end{tabular} &
  \begin{tabular}[c]{@{}l@{}}$\sigma$J\\ mag\end{tabular} &
  \begin{tabular}[c]{@{}l@{}}H\\ mag\end{tabular} &
  \begin{tabular}[c]{@{}l@{}}$\sigma$H\\ mag\end{tabular} &
  \begin{tabular}[c]{@{}l@{}}K$_s$\\ mag\end{tabular} &
  \begin{tabular}[c]{@{}l@{}}$\sigma$K$_s$\\ mag\end{tabular} &
  \begin{tabular}[c]{@{}l@{}}G\\ mag\end{tabular} &
  \begin{tabular}[c]{@{}l@{}}FWHM\\ pixels\end{tabular} &
  $\varepsilon$ &
  \begin{tabular}[c]{@{}l@{}}A$_V$\\ mag\end{tabular} &
  \begin{tabular}[c]{@{}l@{}}A$_J$\\ mag\end{tabular} &
  \begin{tabular}[c]{@{}l@{}}A$_H$\\ mag\end{tabular} &
  \begin{tabular}[c]{@{}l@{}}A$_{ks}$\\ mag\end{tabular} &
  \begin{tabular}[c]{@{}l@{}}A$_G$\\ mag\end{tabular} &
  \begin{tabular}[c]{@{}l@{}}D\\ kpc\end{tabular} \\* \midrule
\endfirsthead
\multicolumn{18}{c}%
{{\bfseries Table \thetable\ continued from previous page}} \\
\toprule
ID &
  \begin{tabular}[c]{@{}l@{}}RA\\ DD\end{tabular} &
  \begin{tabular}[c]{@{}l@{}}DEC\\ DD\end{tabular} &
  \begin{tabular}[c]{@{}l@{}}J\\ mag\end{tabular} &
  \begin{tabular}[c]{@{}l@{}}$\sigma$J\\ mag\end{tabular} &
  \begin{tabular}[c]{@{}l@{}}H\\ mag\end{tabular} &
  \begin{tabular}[c]{@{}l@{}}$\sigma$H\\ mag\end{tabular} &
  \begin{tabular}[c]{@{}l@{}}K$_s$\\ mag\end{tabular} &
  \begin{tabular}[c]{@{}l@{}}$\sigma$K$_s$\\ mag\end{tabular} &
  \begin{tabular}[c]{@{}l@{}}G\\ mag\end{tabular} &
  \begin{tabular}[c]{@{}l@{}}FWHM\\ pixels\end{tabular} &
  $\varepsilon$ &
  \begin{tabular}[c]{@{}l@{}}A$_V$\\ mag\end{tabular} &
  \begin{tabular}[c]{@{}l@{}}A$_J$\\ mag\end{tabular} &
  \begin{tabular}[c]{@{}l@{}}A$_H$\\ mag\end{tabular} &
  \begin{tabular}[c]{@{}l@{}}A$_{ks}$\\ mag\end{tabular} &
  \begin{tabular}[c]{@{}l@{}}A$_G$\\ mag\end{tabular} &
  \begin{tabular}[c]{@{}l@{}}D\\ kpc\end{tabular} \\* \midrule
\endhead
\bottomrule
\endfoot
\endlastfoot
0  & 209.646  & -65.2543 & 16.55 & 0.02 & 16.29 & 0.04 & 16.28 & 0.08 & 17.28 & 3.81 & 0.26 & 2.14 & 0.6  & 0.39 & 0.25 & 1.84 & 106 \\
1  & 209.9556 & -65.7977 & 16.41 & 0.02 & 16.04 & 0.03 & 15.9  & 0.06 & 17.46 & 2.49 & 0.11 & 1.87 & 0.52 & 0.34 & 0.22 & 1.6  & 101 \\
2  & 210.0543 & -65.896  & 15.65 & 0.02 & 15.28 & 0.02 & 15.31 & 0.04 & 16.57 & 3.11 & 0.1  & 1.82 & 0.51 & 0.34 & 0.22 & 1.57 & 101 \\
3  & 210.702  & -65.2801 & 17.23 & 0.03 & 16.91 & 0.06 & 16.91 & 0.1  & 18.28 & 2.64 & 0.21 & 2.34 & 0.66 & 0.43 & 0.28 & 2.02 & 76  \\
4  & 210.8203 & -65.5881 & 17.16 & 0.02 & 16.56 & 0.04 & 16.4  & 0.07 & 19.18 & 2.97 & 0.18 & 1.77 & 0.49 & 0.32 & 0.21 & 1.52 & 74  \\
5  & 211.1837 & -65.2004 & 15.03 & 0.01 & 14.78 & 0.01 & 14.74 & 0.02 & 15.72 & 3.16 & 0.09 & 2.17 & 0.61 & 0.4  & 0.26 & 1.86 & 62  \\
6  & 211.1937 & -65.1237 & 16.37 & 0.02 & 16.01 & 0.03 & 16.09 & 0.06 & 17.42 & 3.66 & 0.24 & 2.16 & 0.6  & 0.4  & 0.25 & 1.85 & 63  \\
7  & 211.5559 & -65.7712 & 15.76 & 0.01 & 15.24 & 0.01 & 15.17 & 0.02 & 17.22 & 2.67 & 0.05 & 1.47 & 0.41 & 0.27 & 0.17 & 1.27 & 59  \\
8  & 211.9328 & -65.6819 & 15.72 & 0.01 & 15.47 & 0.01 & 15.46 & 0.02 & 16.53 & 2.61 & 0.08 & 1.58 & 0.44 & 0.29 & 0.19 & 1.36 & 46  \\
9  & 212.7034 & -65.4799 & 16.05 & 0.01 & 15.68 & 0.02 & 15.56 & 0.04 & 17.1  & 3.5  & 0.3  & 1.77 & 0.5  & 0.33 & 0.21 & 1.52 & 20  \\
10 & 209.6601 & -65.1488 & 15.73 & 0.01 & 15.52 & 0.02 & 15.5  & 0.04 & 16.14 & 3.01 & 0.18 & 2.52 & 0.71 & 0.46 & 0.3  & 2.17 & 107 \\
11 & 209.9389 & -65.7813 & 16.03 & 0.02 & 15.79 & 0.02 & 15.86 & 0.05 & 16.75 & 3.74 & 0.09 & 1.89 & 0.53 & 0.35 & 0.22 & 1.62 & 102 \\
12 & 211.0332 & -65.9285 & 16.26 & 0.01 & 15.87 & 0.02 & 15.73 & 0.05 & 17.42 & 3.68 & 0.16 & 1.34 & 0.38 & 0.25 & 0.16 & 1.15 & 77  \\
13 & 210.2055 & -65.0511 & 15.44 & 0.01 & 15.16 & 0.02 & 15.12 & 0.04 & 15.89 & 2.98 & 0.04 & 2.98 & 0.83 & 0.55 & 0.35 & 2.56 & 93  \\
14 & 211.058  & -64.9099 & 16.42 & 0.02 & 15.82 & 0.03 & 15.86 & 0.04 & 17.19 & 3.87 & 0.15 & 2.13 & 0.6  & 0.39 & 0.25 & 1.83 & 72  \\
15 & 210.0427 & -65.1478 & 15.85 & 0.01 & 15.4  & 0.02 & 15.37 & 0.03 & 17.24 & 2.8  & 0.14 & 2.48 & 0.69 & 0.46 & 0.29 & 2.13 & 96  \\
16 & 209.8827 & -65.3385 & 15.14 & 0.01 & 14.92 & 0.01 & 14.95 & 0.03 & 15.81 & 2.81 & 0.1  & 1.89 & 0.53 & 0.35 & 0.22 & 1.62 & 99  \\
17 & 210.1386 & -65.6147 & 16.46 & 0.03 & 16.0  & 0.04 & 16.02 & 0.08 & 17.92 & 3.15 & 0.03 & 1.77 & 0.5  & 0.33 & 0.21 & 1.52 & 93  \\
18 & 210.6621 & -64.9864 & 15.59 & 0.01 & 15.23 & 0.01 & 15.27 & 0.03 & 16.03 & 2.87 & 0.12 & 2.79 & 0.78 & 0.51 & 0.33 & 2.4  & 81  \\
19 & 210.9808 & -65.0815 & 15.6  & 0.01 & 15.29 & 0.02 & 15.28 & 0.03 & 16.34 & 3.21 & 0.13 & 2.38 & 0.67 & 0.44 & 0.28 & 2.04 & 70  \\
20 & 210.8899 & -65.2929 & 15.44 & 0.02 & 15.14 & 0.02 & 15.19 & 0.03 & 16.38 & 3.12 & 0.02 & 2.15 & 0.6  & 0.4  & 0.25 & 1.85 & 70  \\
21 & 211.0256 & -65.3646 & 15.85 & 0.02 & 15.58 & 0.02 & 15.53 & 0.04 & 16.6  & 4.0  & 0.32 & 2.04 & 0.57 & 0.38 & 0.24 & 1.76 & 66  \\
22 & 211.2093 & -65.5939 & 16.56 & 0.02 & 16.25 & 0.03 & 16.22 & 0.06 & 17.58 & 3.27 & 0.08 & 1.63 & 0.46 & 0.3  & 0.19 & 1.4  & 63  \\
23 & 211.3385 & -65.5525 & 15.5  & 0.02 & 15.17 & 0.02 & 15.16 & 0.03 & 16.27 & 3.99 & 0.21 & 1.66 & 0.46 & 0.31 & 0.2  & 1.43 & 59  \\
24 & 211.9026 & -65.095  & 15.58 & 0.02 & 15.22 & 0.02 & 15.38 & 0.03 & 16.45 & 3.6  & 0.23 & 1.96 & 0.55 & 0.36 & 0.23 & 1.68 & 44  \\
25 & 212.1071 & -65.4834 & 16.24 & 0.02 & 15.89 & 0.02 & 15.89 & 0.05 & 17.03 & 3.18 & 0.21 & 1.6  & 0.45 & 0.29 & 0.19 & 1.37 & 36  \\
26 & 212.6931 & -65.103  & 16.18 & 0.01 & 15.79 & 0.02 & 15.76 & 0.05 & 17.12 & 3.26 & 0.29 & 2.18 & 0.61 & 0.4  & 0.26 & 1.87 & 24  \\
27 & 212.6585 & -65.355  & 15.23 & 0.01 & 14.99 & 0.01 & 14.95 & 0.02 & 15.8  & 3.4  & 0.18 & 1.99 & 0.56 & 0.37 & 0.23 & 1.71 & 18  \\
28 & 212.7143 & -65.4207 & 16.24 & 0.01 & 15.73 & 0.02 & 15.59 & 0.04 & 17.77 & 2.74 & 0.21 & 1.84 & 0.52 & 0.34 & 0.22 & 1.58 & 18  \\
29 & 212.6281 & -65.2966 & 16.58 & 0.02 & 16.14 & 0.02 & 16.0  & 0.05 & 18.1  & 3.4  & 0.32 & 2.05 & 0.57 & 0.38 & 0.24 & 1.76 & 20  \\
30 & 212.8148 & -65.5737 & 16.59 & 0.02 & 16.2  & 0.03 & 16.15 & 0.06 & 17.84 & 2.98 & 0.12 & 1.63 & 0.46 & 0.3  & 0.19 & 1.4  & 21  \\
31 & 211.0072 & -65.0896 & 16.94 & 0.02 & 16.53 & 0.05 & 16.69 & 0.08 & 17.89 & 3.73 & 0.12 & 2.38 & 0.67 & 0.44 & 0.28 & 2.04 & 69  \\
32 & 210.9581 & -65.4085 & 16.61 & 0.02 & 15.99 & 0.03 & 16.04 & 0.06 & 18.22 & 3.56 & 0.14 & 1.99 & 0.56 & 0.37 & 0.23 & 1.71 & 68  \\
33 & 211.7361 & -65.0835 & 16.09 & 0.01 & 15.66 & 0.02 & 15.61 & 0.04 & 17.78 & 3.1  & 0.03 & 1.95 & 0.55 & 0.36 & 0.23 & 1.67 & 49  \\
34 & 211.8109 & -65.0526 & 16.31 & 0.02 & 15.93 & 0.03 & 15.94 & 0.04 & 17.53 & 3.46 & 0.17 & 2.0  & 0.56 & 0.37 & 0.24 & 1.72 & 48  \\
35 & 211.7861 & -65.3681 & 17.46 & 0.03 & 17.06 & 0.05 & 16.99 & 0.13 & 19.43 & 2.99 & 0.26 & 1.56 & 0.44 & 0.29 & 0.18 & 1.34 & 44  \\
36 & 212.0291 & -65.3367 & 16.52 & 0.01 & 16.01 & 0.02 & 15.94 & 0.05 & 17.68 & 3.78 & 0.34 & 1.63 & 0.46 & 0.3  & 0.19 & 1.4  & 37  \\
37 & 211.9312 & -65.6224 & 15.1  & 0.01 & 14.79 & 0.01 & 14.75 & 0.02 & 16.02 & 2.99 & 0.18 & 1.64 & 0.46 & 0.3  & 0.19 & 1.41 & 44  \\
38 & 212.553  & -64.957  & 15.63 & 0.01 & 15.37 & 0.02 & 15.34 & 0.03 & 16.35 & 3.68 & 0.09 & 2.69 & 0.75 & 0.49 & 0.32 & 2.31 & 34  \\
39 & 213.251  & -65.7362 & 16.71 & 0.02 & 16.11 & 0.03 & 15.98 & 0.05 & 18.15 & 2.95 & 0.17 & 1.59 & 0.44 & 0.29 & 0.19 & 1.36 & 28  \\
40 & 213.3396 & -65.7953 & 17.55 & 0.04 & 17.21 & 0.06 & 16.88 & 0.13 & 19.48 & 3.0  & 0.2  & 1.61 & 0.45 & 0.3  & 0.19 & 1.39 & 32  \\
41 & 212.9437 & -65.2087 & 17.07 & 0.03 & 16.58 & 0.03 & 16.39 & 0.1  & 18.49 & 2.88 & 0.18 & 2.01 & 0.56 & 0.37 & 0.24 & 1.73 & 14  \\
42 & 213.0214 & -65.2235 & 15.46 & 0.01 & 15.21 & 0.01 & 15.21 & 0.03 & 16.01 & 3.5  & 0.37 & 2.05 & 0.57 & 0.38 & 0.24 & 1.76 & 11  \\
43 & 213.1648 & -65.1713 & 16.32 & 0.01 & 15.87 & 0.02 & 15.84 & 0.06 & 17.47 & 2.36 & 0.23 & 1.88 & 0.53 & 0.35 & 0.22 & 1.62 & 12  \\
44 & 213.4063 & -65.5202 & 16.73 & 0.02 & 16.4  & 0.02 & 16.3  & 0.07 & 18.05 & 3.6  & 0.08 & 1.63 & 0.46 & 0.3  & 0.19 & 1.4  & 13  \\
45 & 213.6302 & -65.7496 & 16.42 & 0.01 & 15.9  & 0.01 & 15.84 & 0.06 & 18.48 & 3.6  & 0.1  & 1.51 & 0.42 & 0.28 & 0.18 & 1.3  & 30  \\
46 & 213.6953 & -65.6493 & 17.2  & 0.03 & 16.83 & 0.05 & 16.67 & 0.1  & 19.0  & 3.01 & 0.25 & 1.5  & 0.42 & 0.28 & 0.18 & 1.29 & 25  \\
47 & 213.9453 & -64.84   & 16.76 & 0.02 & 16.24 & 0.02 & 16.17 & 0.07 & 17.99 & 3.1  & 0.13 & 2.85 & 0.8  & 0.52 & 0.34 & 2.45 & 40  \\
48 & 214.2796 & -65.4149 & 16.88 & 0.02 & 16.46 & 0.03 & 16.38 & 0.08 & 18.09 & 2.87 & 0.24 & 1.88 & 0.53 & 0.35 & 0.22 & 1.61 & 29  \\
49 & 214.5096 & -64.6824 & 16.9  & 0.02 & 16.38 & 0.02 & 16.64 & 0.08 & 18.02 & 3.85 & 0.13 & 2.56 & 0.72 & 0.47 & 0.3  & 2.2  & 58  \\
50 & 214.7067 & -64.7219 & 16.38 & 0.02 & 16.0  & 0.02 & 16.01 & 0.06 & 17.5  & 2.47 & 0.02 & 2.54 & 0.71 & 0.47 & 0.3  & 2.18 & 60  \\
51 & 214.6406 & -64.5995 & 16.14 & 0.01 & 15.88 & 0.02 & 15.95 & 0.05 & 16.86 & 3.96 & 0.36 & 2.49 & 0.7  & 0.46 & 0.29 & 2.14 & 65  \\
52 & 214.7177 & -64.8725 & 16.78 & 0.03 & 16.47 & 0.03 & 16.56 & 0.1  & 17.63 & 2.92 & 0.22 & 2.68 & 0.75 & 0.49 & 0.32 & 2.3  & 53  \\
53 & 214.9468 & -64.8944 & 16.39 & 0.03 & 16.0  & 0.03 & 15.91 & 0.06 & 17.6  & 2.42 & 0.16 & 2.78 & 0.78 & 0.51 & 0.33 & 2.39 & 58  \\
54 & 215.0736 & -65.0166 & 15.21 & 0.01 & 15.0  & 0.01 & 14.98 & 0.03 & 15.78 & 2.51 & 0.28 & 2.69 & 0.75 & 0.49 & 0.32 & 2.31 & 57  \\
55 & 214.9383 & -65.2082 & 15.02 & 0.01 & 14.8  & 0.01 & 14.8  & 0.02 & 15.54 & 2.31 & 0.05 & 2.13 & 0.6  & 0.39 & 0.25 & 1.83 & 49  \\
56 & 215.42   & -65.4863 & 16.08 & 0.01 & 15.83 & 0.02 & 15.72 & 0.04 & 16.92 & 3.25 & 0.24 & 1.68 & 0.47 & 0.31 & 0.2  & 1.44 & 63  \\
57 & 215.261  & -64.5335 & 17.09 & 0.03 & 16.53 & 0.03 & 16.44 & 0.08 & 18.32 & 3.02 & 0.2  & 2.69 & 0.75 & 0.5  & 0.32 & 2.32 & 81  \\
58 & 213.3722 & -65.621  & 15.56 & 0.01 & 15.31 & 0.02 & 15.24 & 0.03 & 16.25 & 3.63 & 0.21 & 1.58 & 0.44 & 0.29 & 0.19 & 1.36 & 20  \\
59 & 215.677  & -64.7121 & 15.17 & 0.02 & 14.98 & 0.03 & 15.0  & 0.03 & 16.03 & 2.67 & 0.08 & 2.75 & 0.77 & 0.51 & 0.32 & 2.37 & 83  \\
60 & 214.187  & -64.497  & 15.95 & 0.01 & 15.68 & 0.02 & 15.76 & 0.05 & 16.46 & 2.5  & 0.17 & 2.96 & 0.83 & 0.54 & 0.35 & 2.55 & 64  \\
61 & 214.4463 & -64.888  & 16.27 & 0.01 & 15.82 & 0.02 & 15.8  & 0.04 & 17.4  & 3.52 & 0.36 & 2.57 & 0.72 & 0.47 & 0.3  & 2.21 & 46  \\
62 & 214.6598 & -65.1219 & 16.12 & 0.01 & 15.8  & 0.02 & 15.76 & 0.04 & 16.93 & 2.38 & 0.09 & 2.3  & 0.64 & 0.42 & 0.27 & 1.98 & 43  \\
63 & 214.676  & -65.1239 & 16.17 & 0.02 & 15.69 & 0.02 & 15.42 & 0.06 & 17.25 & 3.68 & 0.39 & 2.3  & 0.64 & 0.42 & 0.27 & 1.98 & 43  \\
64 & 214.9766 & -64.3485 & 15.03 & 0.02 & 14.91 & 0.02 & 14.86 & 0.03 & 15.38 & 2.77 & 0.23 & 2.85 & 0.8  & 0.52 & 0.34 & 2.45 & 85  \\
65 & 215.9346 & -64.6718 & 16.63 & 0.02 & 16.09 & 0.02 & 15.95 & 0.04 & 18.11 & 3.96 & 0.3  & 2.41 & 0.67 & 0.44 & 0.28 & 2.07 & 91  \\
66 & 215.9726 & -64.5859 & 15.02 & 0.01 & 14.9  & 0.01 & 14.86 & 0.02 & 15.24 & 2.93 & 0.22 & 2.36 & 0.66 & 0.43 & 0.28 & 2.03 & 95  \\
67 & 216.1489 & -64.9173 & 17.68 & 0.05 & 17.4  & 0.07 & 17.19 & 0.15 & 18.85 & 3.44 & 0.05 & 1.9  & 0.53 & 0.35 & 0.22 & 1.63 & 89  \\
68 & 213.6556 & -65.0388 & 16.91 & 0.02 & 16.46 & 0.03 & 16.17 & 0.08 & 17.81 & 3.02 & 0.2  & 2.33 & 0.65 & 0.43 & 0.27 & 2.0  & 24  \\
69 & 214.9895 & -65.5057 & 16.78 & 0.02 & 16.34 & 0.02 & 16.19 & 0.06 & 17.85 & 2.91 & 0.17 & 1.76 & 0.49 & 0.32 & 0.21 & 1.51 & 51  \\
70 & 214.9922 & -64.8638 & 16.82 & 0.02 & 16.54 & 0.03 & 16.56 & 0.09 & 17.55 & 2.32 & 0.24 & 2.73 & 0.77 & 0.5  & 0.32 & 2.35 & 60  \\
71 & 215.2915 & -65.1353 & 16.8  & 0.02 & 16.48 & 0.02 & 16.55 & 0.08 & 17.52 & 3.71 & 0.25 & 2.14 & 0.6  & 0.39 & 0.25 & 1.84 & 60  \\
72 & 215.554  & -65.0999 & 15.74 & 0.01 & 15.44 & 0.02 & 15.4  & 0.03 & 16.53 & 2.33 & 0.03 & 2.1  & 0.59 & 0.39 & 0.25 & 1.8  & 68  \\
73 & 215.7254 & -64.4617 & 16.35 & 0.02 & 15.98 & 0.02 & 16.05 & 0.06 & 17.16 & 2.5  & 0.18 & 2.58 & 0.72 & 0.48 & 0.3  & 2.22 & 95  \\
74 & 216.3452 & -65.0076 & 16.21 & 0.01 & 15.82 & 0.01 & 15.66 & 0.04 & 17.3  & 3.22 & 0.3  & 1.7  & 0.48 & 0.31 & 0.2  & 1.46 & 92  \\
75 & 215.7889 & -65.4676 & 17.89 & 0.03 & 17.19 & 0.03 & 16.78 & 0.11 & 19.73 & 2.66 & 0.07 & 1.42 & 0.4  & 0.26 & 0.17 & 1.22 & 73  \\
76 & 215.8472 & -65.4723 & 15.94 & 0.01 & 15.61 & 0.01 & 15.41 & 0.03 & 16.95 & 3.58 & 0.34 & 1.39 & 0.39 & 0.26 & 0.16 & 1.19 & 75  \\
77 & 214.4796 & -65.3112 & 15.08 & 0.04 & 14.89 & 0.05 & 14.53 & 0.03 & 15.66 & 2.77 & 0.04 & 2.01 & 0.56 & 0.37 & 0.24 & 1.73 & 35  \\* \bottomrule
\end{longtable}
\twocolumn
\end{landscape}
\end{appendix}
\end{document}